\documentclass[fleqn,usenatbib]{mnras}

\usepackage[T1]{fontenc}
\usepackage{multirow}
\usepackage{float}

\DeclareRobustCommand{\VAN}[3]{#2}
\let\VANthebibliography\thebibliography
\def\thebibliography{\DeclareRobustCommand{\VAN}[3]{##3}\VANthebibliography}

\usepackage{graphicx}	
\usepackage{amsmath}	
\usepackage{amssymb}	

\title[YETI follow-up observations of CVSO\,30]{YETI follow-up observations of the T Tauri star CVSO\,30 with transit-like dips}

\author[R. Bischoff et al.]{R. Bischoff,$^{1}$\thanks{E-mail: richard.bischoff@uni-jena.de} St. Raetz,$^{2}$ M. Fern\'{a}ndez,$^{3}$ M. Mugrauer,$^{1}$ R. Neuh\"{a}user,$^{1}$ P.\,C. Huang,$^{4}$ \newauthor W.\,P. Chen,$^{4,5}$ A. Sota,$^{3}$ J. Jim\'{e}nez Ortega,$^{3}$ V.\,V. Hambaryan,$^{1,6}$ P. Zieli\'{n}ski,$^{7,8}$ M. Dr\'{o}{\.z}d{\.z},$^{9}$ \newauthor W. Og{\l}oza,$^{9}$ W. Stenglein,$^{1}$ E. Hohmann,$^{1}$ and K.-U. Michel$^{1}$\\
$^{1}$Astrophysical Institute and University Observatory, Friedrich-Schiller-Universit\"{a}t Jena, Schillerg\"{a}{\ss}chen 2, 07745 Jena, Germany\\
$^{2}$Institute for Astronomy and Astrophysics T\"{u}bingen, Eberhard-Karls-Universit\"{a}t T\"{u}bingen, Sand 1, 72076 T\"{u}bingen, Germany\\
$^{3}$Instituto de Astrof\'{\i}sica de Andaluc\'{\i}a, CSIC, Apdo. 3004, E-18080 Granada, Spain\\
$^{4}$Graduate Institute of Astronomy, National Central University, 300 Zhongda Road, Zhongli, Taoyuan 32001, Taiwan \\
$^{5}$Department of Physics, National Central University, 300 Zhongda Road, Zhongli, Taoyuan 32001, Taiwan\\
$^{6}$Byurakan Astrophysical Observatory, Byurakan 0213, Aragatzotn, Armenia\\
$^{7}$Institute of Astronomy, Faculty of Physics, Astronomy and Informatics, Nicolaus Copernicus University in Toru{\'n},\\
\, ul. Grudzi\k{a}dzka 5, 87-100 Toru{\'n}, Poland\\
$^{8}$Astronomical Observatory, University of Warsaw, Al. Ujazdowskie 4, 00-478 Warszawa, Poland \\
$^{9}$Mt. Suhora Observatory, Pedagogical University, ul. Podchorazych 2, 30-084 Krakow, Poland}

\date{Accepted 2022 January 27. Received 2021 January 13; in original form 2021 November 12}

\pubyear{2022}

\begin{document}
\label{firstpage}
\pagerange{\pageref{firstpage}--\pageref{lastpage}}
\maketitle

\begin{abstract}
The T Tauri star CVSO\,30, also known as PTFO\,8-8695, was studied intensively with ground based telescopes as well as with satellites over the last decade. It showed a variable light curve with additional repeating planetary transit-like dips every $\sim10.8$\,h. However, these dimming events changed in depth and duration since their discovery and from autumn 2018 on, they were not even present or near the predicted observing times. As reason for the detected dips and their changes within the complex light curve, e.g. a disintegrating planet, a circumstellar dust clump, stellar spots, possible multiplicity and orbiting clouds at a Keplerian co-rotating radius were discussed and are still under debate. In this paper, we present additional optical monitoring of CVSO\,30 with the meter class telescopes of the Young Exoplanet Transit Initiative in Asia and Europe over the last seven years and characterize CVSO\,30 with the new Early Data Release 3 of the ESA \textit{Gaia} Mission. As a result, we describe the evolution of the dimming events in the optical wavelength range since 2014 and present explanatory approaches for the observed variabilities. We conclude that orbiting clouds of gas at a Keplerian co-rotating radius are the most promising scenario to explain most changes in CVSO\,30's light curve.
\end{abstract}


\section{Introduction}

While most exoplanets are Gyr old, including in particular transiting planets, it would be best to study planet formation and early evolution with young planets (age $<100$\,Myr). Therefore, the Young Exoplanet Transit Initiative \citep[YETI;][]{neuhaeuser} searched for planet candidates in young open stellar clusters.
One of our first targets was the 25\,Ori cluster, which was observed within the YETI network since 2010. As a result, we could confirm the detection of a transit-like signal within the light curve of CVSO\,30, which was first found by \cite{vaneyken}. \cite{schmidt} discovered an additional wide companion planet candidate via direct imaging.
CVSO\,30, which is also known as PTFO\,8-8695, is a weak-line T Tauri star with spectral type M3 \citep{briceno}, which showed initially brightness dips of $\sim 35$\,mmag every $\sim10.8$\,h lasting about $100$\,min. However, shape, depth and duration of these dips changed significantly over time as reported by \cite{vaneyken} and \cite{raetz}. Furthermore, the dimming events can show either no wavelength dependence \citep{raetz} or, in other epochs, depths that decrease at larger wavelengths \citep{yu,onitsuka,tanimoto}. On the one hand, CVSO\,30 seems to show no dimming at some epochs \citep{koen2015} and on the other hand, multiple dips per period were detected. \cite{tanimoto} monitored CVSO\,30 intensively in the $I$-band and in $JHK_{s}$-filters between 2014 and 2018. They found that CVSO\,30 showed three different fading events, namely "dip-A" appeared $0.1$ earlier in phase compared to the dimming reported in \cite{vaneyken}, which they called "dip-B". Another "dip-C" was discovered in November 2018, $0.5$ later in phase.
As diverse as the detected changes in CVSO\,30's light curve were also the attempts to explain their origin during the last decade. \cite{vaneyken} argued that the signal was caused by a Jovian planet based on radial velocity $(RV)$ measurements and adaptive optics imaging, where no significant $RV$ variation or background source could be detected. In order to explain the changing depth, duration as well as the dis- and reappearance of the transit-like signal \cite{barnes} proposed a misalignment between the rotational axis of the star and the orbital plane of a precessing planet. This hypothesis agrees with \cite{ciardi}, but the model needed further fine tuning to fit their photometric and spectroscopic data.
In contrast to this, the planetary scenario is considered unlikely by \cite{yu}, because the target does not show the Rossiter-McLaughlin effect or changes in its radial velocity between 2011 and 2013. Also star spots are usually visible for half a rotational period and this contradicts the short duration of the observed fading events. Even if multiple stellar spots were considered, this scenario could reproduce the signal, but it is based on assuming a complex, and stable pattern \citep{yu}.
\cite{tanimoto} suggest the idea that the original fading event in \cite{vaneyken} consists of a periodically combined dust cloud and a precessing planet, which split up in 2014 to explain the detection of multiple dips per period near the predicted observing times. Additionally, the newly found dip-C could be the result of an accretion hotspot or a dust cloud.
Recent publications of \cite{bouma} and \cite{koen2020} showed that the light curve from the Transiting Exoplanet Survey Satellite  \citep[\textit{TESS};][]{ricker} contains two different periods (11.98\,h and 10.76\,h) and that CVSO\,30 can be a binary star with no planetary companion. \cite{bouma} presented also several other possible explanations, which need to be considered further. \cite{koen2021} combined the binary scenario with star spot models to explain the variability of CVSO\,30 based on \textit{TESS} measurements in one passband. However, a filling factor of roughly 0.5 seems by far too large for normal spots and would have been probably detected in previous high-resolution, high signal-to-noise spectra, as e.g. in the case of the weak-line T Tauri star P1724 \cite{neuhaeuser2}.

In this paper, we describe our observations in the optical wavelength range in section\,\ref{observation}. In section\,\ref{datared} we explain the data reduction and the routine for photometric measurements. The following section\,\ref{lcana} presents our light curve analysis. We test scenarios for the cause of CVSO\,30's variability in section\,\ref{expl}. Our results are discussed and we give a conclusion in the final section of this paper.

\section{Observations}\label{observation}

In contrast to the original YETI campaigns, where the 25\,Ori cluster was continuously monitored for 7 to 12 subsequent nights within multiple runs in several years, we focused this time completely on the predicted time slots of the dimming event, according to ephemeris presented in \cite{raetz}. An overview of the involved observatories and instrumentation, which participated within our YETI follow-up observations of CVSO\,30 since autumn 2014, is given in Table\,\ref{tab:observatory}. Each monitoring includes typically about 1\,h of observation time before and after the start/end of the predicted dimming event. The observations were usually carried out in the $R$-band filter with individual image integration times up to a few minutes, as listed in the observation log in Table\,\ref{tab:obslog}, for sufficient photometric precision and time resolution.

After receiving message from T.\,O.\,B.\,Schmidt (priv. communication) in summer 2019, who reported a phase shifted dimming within the \textit{TESS} light curve of CVSO\,30, we extended our monitoring also to these additional time slots. \cite{tanimoto} had found this signal independently within their data in November 2018.

Our photometric follow-up observations of CVSO\,30 have a total integration time of about 164.4\,h spanning over the range of time between October 2014 and February 2021.

\begin{table*}
\centering
\caption{Observatories and instruments of the YETI network, which took part in the follow-up observations.}
\label{tab:observatory}
\resizebox{\linewidth}{!}{
\begin{tabular}{ccccccccc}
\hline
Observatory & Abbrev. 	& Long. (E) & Lat. (N)	& Altitude 	& Mirror $\diameter$ 	 & CCD 				& \# Pixel &  FoV \\
			&			& [deg]		& [deg]		& [m]		& [m]		 &					&				& [arcmin]\\
\hline
Lulin/Taiwan & LOT 			& 120.5 & 23.3	& 2862 		& 1.0	& Apogee U42$^{a}$			& 2048\,x\,2048 & 11.0\,x\,11.0\\
Suhora/Poland& Suhora		& 20.1	& 49.6	& 1009		& 0.6	& Apogee Aspen CG47$^{b}$ 	& 1024\,x\,1024 & 20.0\,x\,20.0\\
Jena/Germany & GSH			& 11.5	& 50.9	& 367		&\,\,\,0.9$^{c}$	& E2V\,CCD42-40$^{d}$	& 2048\,x\,2048 & 52.8\,x\,52.8\\
T\"{u}bingen/Germany& IAAT	& 9.1	& 48.5	& 400		& 0.8	& SBIG ST-L-1001E$^{e}$ 		& 1024\,x\,1024 &	13.6\,x\,13.6\\
Sierra Nevada/Spain & OSN 	& -3.4 	& 37.1	& 2896		& 1.5	& VersArray:2048B$^{f}$		& 2048\,x\,2048 & 7.9\,x\,7.9 \\
Tenerife/Spain   & OGS		& -16.5	& 28.3	& 2393		& 1.0 	& EEV 42-40$^{g}$ 		& 4\,x\,[2048\,x\,2048] & 42.5\,x\,42.5 \\
\hline
\end{tabular} }
\begin{flushleft}
$^{a}$ \cite{Huang2019}; $^{b}$ \cite{siwak2019}; $^{c}$ 0.6\,m in Schmidt mode; $^{d}$ \cite{mugrauer};\\
$^{e}$ \url{http://astro.uni-tuebingen.de/about/teleskop/telkam.shtml}; $^{f}$ \cite{Ortiz2006};\\
$^{g}$ \url{http://research.iac.es/OOCC/iac-managed-telescopes/ogs/} \\
\end{flushleft}
\end{table*}				

\begin{table}
\centering
\caption{Observation log. Summary of all follow-up observations of CVSO\,30 within the YETI network since 2014. For each epoch, we list the associated observing date (start of observations), site, used filter, number of exposures ($N_{\text{exp}}$), and individual detector integration time of each frame (DIT).}
\label{tab:obslog}
\begin{tabular}{ccccc}
\hline
Date  & Observatory & Filter & $N_{\text{exp}}$ & DIT $[\text{s}]$ \\
\hline
2014 Oct 23 & OSN & R	& 172	& 90\\
2014 Oct 28 & OSN & R	& 165	& 90 \\
2014 Nov 02 & GSH & R	& 66	& 180\\
2014 Nov 06 & OSN & R	& 145	& 100\\
2014 Nov 19 & OSN & R	& 62	& 90\\
2014 Dec 11 & OSN & R	& 125	& 100\\
2015 Feb 11 & GSH & R	& 74	& 180\\
&  & 	& 	&\\
2015 Nov 14 & OGS & R	& 60	& 210\\
2015 Nov 19 & OGS & clear	& 103	& 135\\
2015 Nov 23 & OSN & R	& 76	& 180\\
2016 Feb 02 & OSN & R	& 100	& 180\\
&&&&\\
2016 Oct 26 & OSN & R	& 30	& 180\\
2017 Jan 05 & OSN & R	& 69	& 240\\
2017 Jan 14 & OSN & R	& 13	& 240\\
2017 Jan 23 & OSN & R	& 86	& 180\\
&&&&\\
2017 Oct 27 & OSN & R	& 70	& 200\\
2017 Nov 13 & OSN & R	& 84	& 200\\
2017 Nov 22 & OSN & R	& 67	& 200\\
2017 Nov 22 & GSH & R	& 106	& 120\\
2018 Jan 23 & OSN & R	& 123	& 200\\
2018 Feb 14 & IAAT& clear	& 124	& 120\\
&&&&\\
2018 Oct 14 & GSH & R	& 47	& 210\\
2018 Oct 31 & Suhora & V, I	& 80, 80	& 120, 60\\
2018 Nov 08 & LOT & R	& 60	& 120\\
2018 Nov 09 & GSH & R	& 32	& 300\\
2018 Nov 14 & GSH & R	& 50	& 210\\
2018 Dec 01 & OSN & R	& 85	& 180\\
2018 Dec 04 & LOT & R	& 136	& 120\\
2018 Dec 09 & LOT & R	& 47 	& 120\\
2018 Dec 13 & LOT & R	& 80	& 120\\
2018 Dec 17 & LOT & R	& 137	& 120\\
2018 Dec 18 & LOT & R	& 137	& 120\\
2018 Dec 26 & LOT & R	& 37	& 120\\
2018 Dec 27 & LOT & R	& 137	& 120\\
&&&&\\
2019 Oct 31 & GSH & R	& 65	& 180\\
2019 Nov 09 & GSH & R	& 52	& 180\\
2019 Dec 05 & GSH & R	& 78	& 180\\
2020 Jan 15 & GSH & R	& 53	& 180\\
2020 Jan 23 & GSH & R	& 60	& 180\\
2020 Feb 07 & GSH & R   & 38    & 180 \\
&&&&\\
2020 Oct 20 & GSH & R	& 62	& 180\\
2020 Nov 23 & GSH & R	& 77	& 180\\
2020 Nov 25 & GSH & R	& 63	& 180\\
2021 Jan 11 & OSN & R	& 66	& 200\\
2021 Jan 15 & OSN & R	& 79	& 200\\
2021 Jan 29 & OSN & R	& 139	& 90\\
2021 Feb 02 & OSN & R	& 82	& 200\\
\hline
\end{tabular}
\end{table}

\section{Data reduction and photometry}\label{datared}

The data were processed with standard image reduction routines based on IRAF\footnote{IRAF is distributed by the National Optical Astronomy Observatories, which are operated by the Association of Universities for Research and Astronomy, Inc., under cooperative agreement with the National Science Foundation.} \citep{tody}, which include bias, dark and flat-field correction.

With our routine, we can perform photometry on all stars within the field of view simultaneously. Therefore, we create a list of pixel coordinates for all detectable light sources with Source Extractor \citep[SExtractor;][]{bertin}. This list was then used as a reference to remove tracking offsets between the individual images, which was also carried out with SExtractor. We determined the optimal aperture size with IRAF for each night separately, by using 15 different apertures, ranging from one up to two average full width half maxima of point sources, detected in the individual observing nights. The standard deviations of the instrumental magnitude differences, of a subset of the brightest, non-variable stars, were then calculated for all 15 apertures, and we chose as optimal aperture the one, with the smallest sum of standard deviations. The optimized aperture was then utilized for aperture photometry on all stars within the field of view.

At next, we performed differential photometry with the program PHOTOMETRY from \cite{broeg}, which creates an artificial comparison star. This artificial star includes information of all detected stars, but they are weighted depending on their stability during the observation. Variable stars have typically a higher standard deviations during the processed time series and therefore, they are weighted lower than stable stars. For more details on the used photometry routine see \cite{errmann} or \cite{errmanndiss}.

Finally, we get a list for each star within the field of view, which includes the heliocentric Julian date (HJD), the determined relative magnitude and its uncertainty.

\section{Light curve analysis}\label{lcana}

The unprocessed light curves are shown in appendix\,\ref{appLC}. Since CVSO\,30 is a T Tauri star, its light curve is impacted by stellar variability which had to be considered and characterized before further investigation. We treated every light curve individually and detrended them by fitting polynomials of the third order to the out-of-event measurements. This was done for the listed nights in Table\,\ref{tab:transitsETD}, which exhibit sufficient out-of-event observing time. The nights without significant fading events were not included in Table\,\ref{tab:transitsETD}, but they are presented in Fig.\,\ref{fig:20142015}\,-\,\ref{fig:20202021}. The detrended (if applicable) and also the original photometric measurements of all light curves are provided as online supplementary material.

Furthermore, we found within our light curves three flare-like events. Their flux increase was determined by fitting third order polynomials to the unprocessed photometric measurements outside the flare-like event and then comparing the expected flux, given by the polynomial, to the corresponding actual measurement during the flare. The associated results are given in Table\,\ref{flare}.

\begin{table}
\centering
\caption{Properties of the detected flare-like events during our monitoring campaign.}
\label{flare}
\begin{tabular}{ccc}
\hline
Date 		& Observatory & maximal flux increase [\%]  \\
\hline
2018 Jan 23 & OSN & $10.1\pm0.8$  \\
2018 Nov 14 & GSH & $\,\,\,4.7\pm1.0$  \\
2018 Dec 27 & LOT & $\,\,\,3.5\pm0.5$  \\
\hline
\end{tabular}
\end{table}

\subsection{Transit fitting}

The detrended light curves of CVSO\,30 were further analyzed using the "Exoplanet Transit Database" \citep[\texttt{ETD};][]{ETD, ETD2}.

\texttt{ETD} is an on-line portal, which can be utilized to fit synthetic transit light curves to observational data. The website determines mid-time, duration and depth of the fading event by using non-linear least-squares algorithm
and also removing systematic trends by a second-order polynomial \citep{ETD}. As input parameter, a first estimation of the mid-transit time and transit duration are needed together with the radii ratio of host star and companion, as well as the impact parameter and the linear limb darkening coefficient. These variables were obtained from stellar mass, radius and orbital period, as listed in Table\,\ref{tab:stardata}. The linear limb darkening coefficient for the $R$-band filter ($u=0.717\pm0.033$) is estimated, based on the given effective temperature and surface gravity, from the work of \cite{claret}.

The radius of CVSO\,30 was determined by calculating its bolometric magnitude
\begin{equation}
M_{\text{bol}}=M_{G}+BC_{G},
\end{equation}
where $M_{G}$ is the absolute brightness in the $G$-band (see section\,\ref{comp} for details) and $BC_{G}$ the corresponding bolometric correction. We derived $BC_{G}=(-1.445\pm0.072)$\,mag from the website of "MESA Isochrones \& Stellar Tracks\footnote{\url{http://waps.cfa.harvard.edu/MIST/model_grids.html}}", taking into account CVSO\,30's effective temperature, extinction, metallicity and surface gravity (values in Table\,\ref{tab:stardata}). At next, we used
\begin{equation}
M_{\text{bol}}=M_{\text{bol},\sun}-2.5\log\left(\dfrac{L}{L_{\sun}}\right)
\end{equation}
 to calculate CVSO\,30's luminosity \citep[using $M_{\text{bol},\sun}=4.74$\,mag from][]{prsa} and via
\begin{equation}
R\approx  \left(\dfrac{L}{L_{\sun}}\right)^{0.5} \cdot \left(\dfrac{T_{\text{eff},\sun}}{T_{\text{eff}}}\right)^{2} \cdot R_{\sun}
\end{equation}
its radius.
The optimal light curve parameters are found with \texttt{ETD} by iterating the input parameter until the output parameters are consistent with each other, within their one sigma uncertainties, in five consecutive fitting attempts. \texttt{ETD} needs as input parameters the limb darkening, the impact factor, the radii ratio of the planet candidate and the star, as well as a specification of the expected transit center time and duration.
\begin{table}
\centering
\caption{Physical parameters of the CVSO\,30.}
\label{tab:stardata}
\begin{tabular}{llc}
\hline
Parameter 								& Value 						& Ref. \\
\hline
RA (J2000) [h\,:\,m\,:\,s]				& \,\,\,\,\,\,05\,:\,25\,:\,07.6 			& [1] \\
Dec (J2000) [deg\,:\,m\,:\,s]			& +\,01\,:\,34\,:\,24.5			& [1] \\
Mass  $M$ [M$_{\sun}$]		& \,\,\,$0.502\pm0.038$ 				& [2] \\
Radius $R$ [R$_{\sun}$]		& \,\,\,\,\,\,$1.69\pm0.16$				& this work \\
Effective temperature $T_{\text{eff}}$ [K]& \,\,\,\,\,$3448\,\,_{-12}^{+43}$ 			& [2] \\
Surface gravity $\log(g)$ [dex]	& \,\,\,\,\,\,$3.84\,\,_{-0.04}^{+0.02}$		& [2] \\
Metallicity $[\text{Fe/H}]$ [dex]				&\,\,\,$0.500\pm0.001$				& [2] \\
Distance $d$ [pc]						& \,\,\,\,\,\,\,\,$334\,\,_{-3}^{+4}$				& [3] \\
Age [Myr] 								& \,\,\,\,\,\,\,\,\,$8.5\pm1.2$ 					& [4] \\
Apparent brightness $m_{\text{G}}$ [mag] 		& $15.101\pm0.003$				& [5]\\
Extinction $A_{\text{G}}$	[mag]				& \,\,\,$0.195\pm0.049$				& this work \\
Absolute brightness $M_{\text{G}}$ [mag] 		& \,\,\,$7.286\,\,_{-0.078}^{+0.075}$		& this work\\
\hline
\end{tabular}
\begin{flushleft}
[1] \cite {briceno}, [2] \cite {queiroz}, [3] \cite{bailer-jones}, [4] \cite{kounkel2018}, [5] \cite{gaiaedr3}
\end{flushleft}
\end{table}
The best-fitting parameters for all detected dimming events are listed in Table\,\ref{tab:transitsETD}. We use the designation of \cite{tanimoto} to distinguish the three phase-shifted fading events of CVSO\,30. The transit center times of the detected dimming events were converted from HJD$_{\text{UTC}}$ into $\text{BJD}_{\text{TDB}}$ with the online converter\footnote{\url{https://astroutils.astronomy.osu.edu/time/hjd2bjd.html}}, based on \cite{eastman}.

\subsection{Investigation of the three phase shifted dimming events}

\begin{figure}
	\includegraphics[width=\columnwidth]{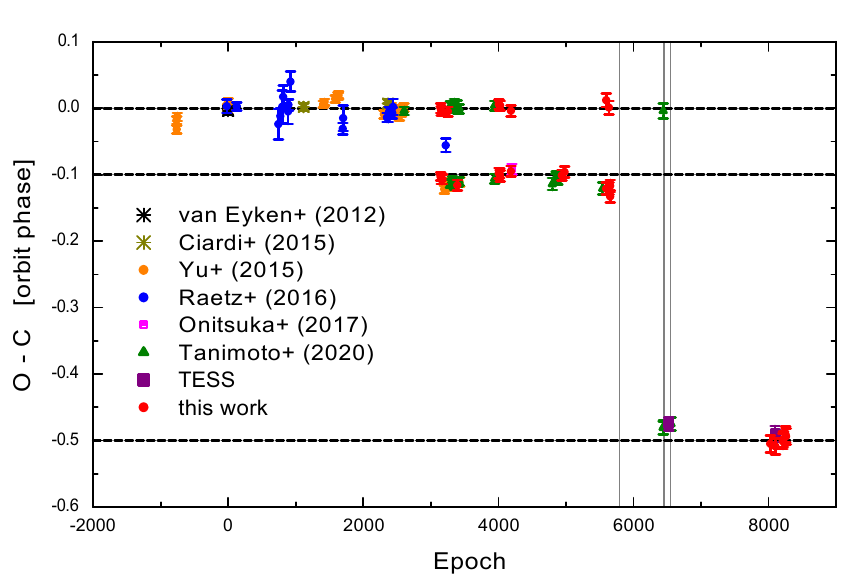}
	\caption{The O-C diagram of CVSO\,30. Only complete fading events from \protect\cite{raetz} are plotted. The epochs with detected flare-like events are illustrated as gray vertical lines.}
	\label{fig:OC}
\end{figure}

\begin{figure*}
	\includegraphics[width=15cm]{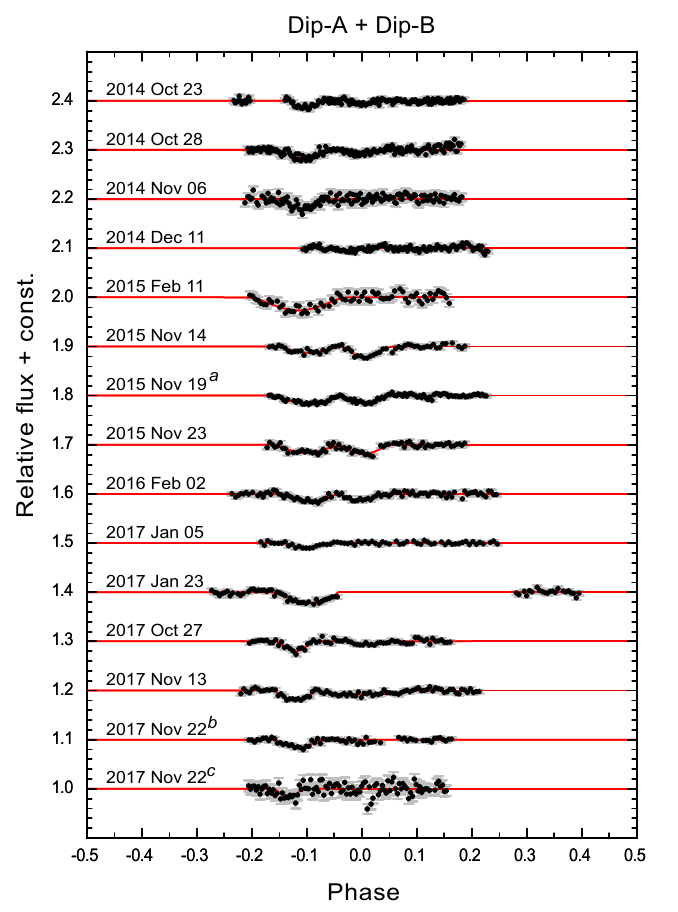}
	\caption{Detrended and phase folded $R$-band light curves of dip-A and dip-B according to the ephemeris in Eqn.\,\ref{ephem}. \newline $^{a}$ clear filter, $^{b}$ OSN, $^{c}$ GSH}
	\label{fig:phaseAB}
\end{figure*}

\begin{figure*}
	\includegraphics[width=15cm]{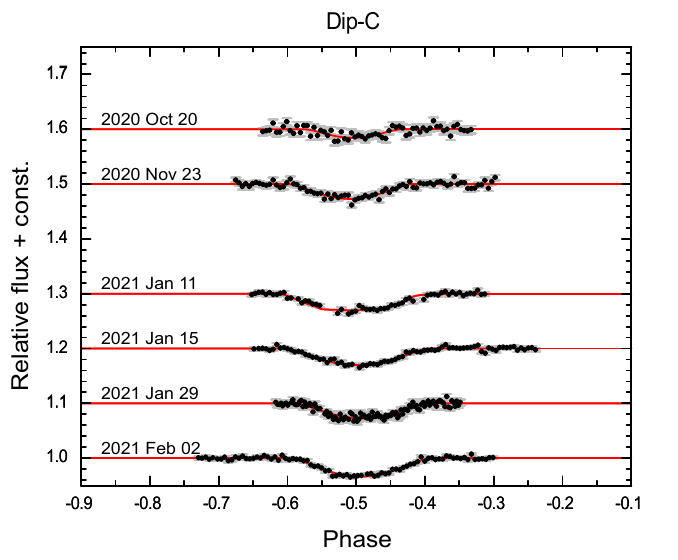}
	\caption{Detrended and phase folded $R$-band light curves of dip-C according to the ephemeris in Eqn.\,\ref{ephem}.}
	\label{fig:phaseC}
\end{figure*}

We found three different dimming events within our observations, which we call dip-A, dip-B and dip-C based on the work of \cite{tanimoto}. Furthermore, their ephemeris
\begin{equation}
\label{ephem}
T_{0}[\text{BJD}_{\text{TDB}}]=2455543.943\pm0.002,
\end{equation}
\vspace*{-6mm}
\begin{equation*}
P[\text{d}]=0.4483993\pm0.0000006,
\end{equation*}
work excellent to predict the occurrence of dip-B within our data, as shown in appendix\,\ref{appLC} and also to characterize the temporal occurrence of dip-A and dip-C. We show the "observed minus calculated" (O-C) diagram in Fig.\,\ref{fig:OC} for our detected fading events together with data from \cite{vaneyken}, \cite{ciardi}, \cite{yu}, \cite{raetz}, \cite{onitsuka}, \cite{tanimoto} and \textit{TESS}, spanning over a decade of observations. Each dip considered separately shows no significant phase shift within its $3\,\sigma$ uncertainties. Therefore, further updates on their ephemeris are not necessary. The detrended and phase folded light curves with the best-fitting models are shown in Fig.\,\ref{fig:phaseAB} and \ref{fig:phaseC}.


\begin{figure*}
	\includegraphics[width=\textwidth]{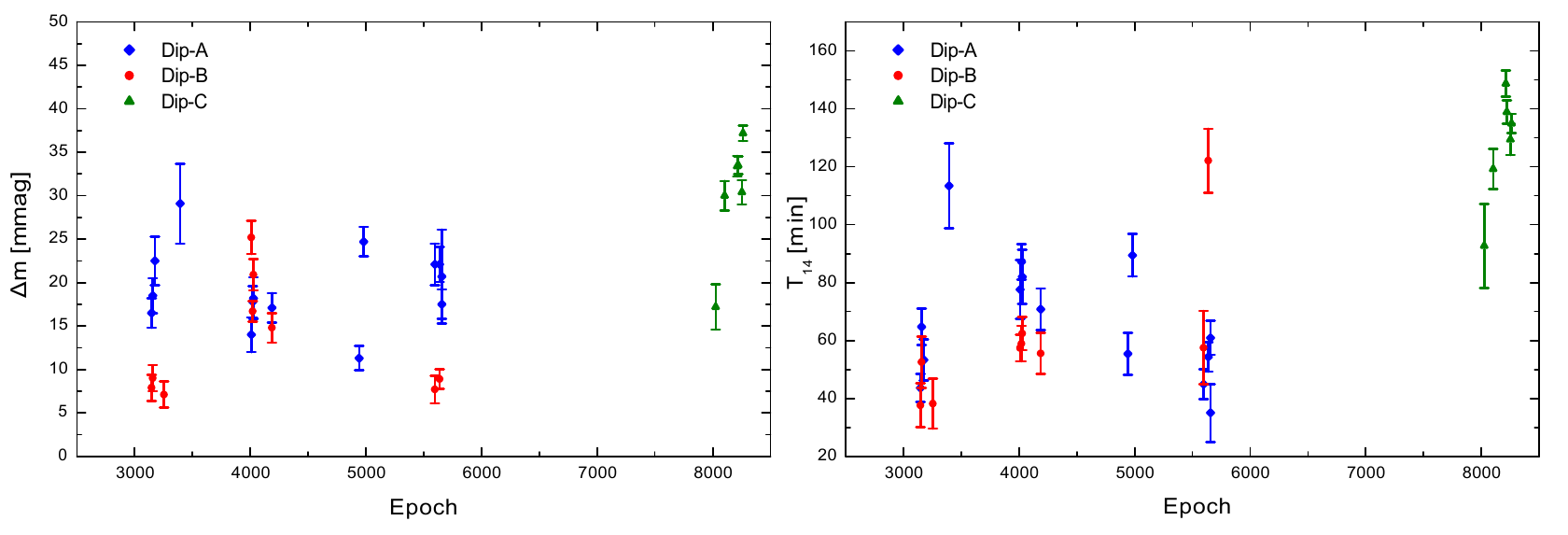}
	\caption{Chronological development of the depths (left) and durations (right) for the phase shifted fading events of CVSO\,30 as measured within our follow-up observations in the $R$-band.}
	\label{fig:dip}
\end{figure*}

As expected, dip-B occurs on average at phase $0.002\pm0.005$ for our observations, using its ephemeris, given in Eqn.\,\ref{ephem}. Dip-A was visible at an average orbit phase of $-0.108\pm0.012$ and dip-C at $-0.500\pm0.007$ during our monitoring campaign, in comparison to dip-B. We have found flare-like events within CVSO\,30's light curve and their temporal appearances are indicated as gray vertical lines in Fig.\,\ref{fig:OC}. The strongest flare-like event took place after the last significant detection of dip-A and dip-B, and before the first appearance of dip-C.
The chronological development of the depths and durations for the three different dimming events are shown in Fig.\,\ref{fig:dip}. The average depth of dip-A $\Delta m_{\text{A}}=(19.4\pm4.5)$\,mmag indicates that dip-A has a comparable depth to dip-B with $\Delta m_{\text{B}}=(13.1\pm6.6)$\,mmag over the whole monitoring campaign. However, dip-B seems to be significantly deeper in season 2015/2016 with $\Delta m_{\text{B}15/16}=(19.4\pm4.6)$\,mmag in comparison to the detections in all other seasons $\Delta m_{\text{B\,other}}=(8.1\pm0.8)$\,mmag. Both, dip-A and dip-B, were not observable anymore since autumn 2017, as illustrated in Fig.\,\ref{fig:20182019} to \ref{fig:20202021}. This contradicts the report in \cite{tanimoto}, namely that they have detected dip-B on November 09 in 2018. This mentioned detection is also consistent with noise. Dip-C has an average depth of $\Delta m_{\text{C}}=(30.3\pm6.9)$\,mmag.

During our follow-up observations dip-A shows an average duration of $T_{14\,\text{A}}=(66.6\pm21.5)$\,min, while dip-B and dip-C last on average $T_{14\,\text{B}}=(60.2\pm24.8)$\,min and $T_{14\,\text{C}}=(127.3\pm19.6)$\,min.

\section{Explanatory approaches for the observed variabilities}\label{expl}

\subsection{Comparison with Tanimoto et al. (2020)}

In this subsection we are following the explanation approaches of \cite{tanimoto}, who give four possible explanations for the dimming events, namely
\begin{enumerate}
\item a cool star spot,
\item an accretion hotspot,
\item a Jovian planet,
\item a circumstellar dust clump,
\end{enumerate}
and test them in the context of their photometric measurements in the near infrared. We will investigate all these scenarios for the individual dimming events, which were not ruled out already by \cite{tanimoto}, based on our observations. \newline

A circumstellar dust clump, consisting of an opaque core and an optically thin dust halo, was the only remaining cause for dip-A that was not falsified. The observed fading of the flux $\delta_{\text{obs}}(\lambda)$ at a particular wavelength is described by
\begin{equation}\label{Tiefe}
\delta_{\text{obs}}(\lambda)=f_{\text{core}}+f_{\text{halo}} \tau_{\text{V}} \left[ a(\lambda^{-1})+\frac{b(\lambda^{-1})}{R_{\text{V}}}\right],
\end{equation}
with the filling factor of the core $f_{\text{core}}$ and the dust halo $f_{\text{halo}}$. The depth in the $V$-band is given by $\tau_{\text{V}}$, and $a(\lambda^{-1})$ and $b(\lambda^{-1})$ are wavelength-dependent coefficients, as defined by \cite{car}. The ratio
\begin{align*}
R_{\text{V}}=\frac{A_{\text{V}}}{E(B-V)},
\end{align*}
was determined by \cite{tanimoto} to be $R_{\text{V}}=5.3$ for dip-A in both, season 2014 and 2016, based on its wavelength dependence. The corresponding best-fitting values for season 2014 are $f_{\text{core}}=0.01$, $f_{\text{halo}} \tau_{\text{V}}=0.014$ and for season 2016 $f_{\text{core}}=0.005$, $f_{\text{halo}} \tau_{\text{V}}=0.0135$. Given these constraints from \cite{tanimoto}, we found that the typical $R$-band depths of dip-A for the corresponding seasons, namely $\delta_{\text{dip-A}}=0.020\pm0.005$ and $\delta_{\text{dip-A}}=0.016\pm0.009$, from our measurements fulfill Eqn.\,\ref{Tiefe} within their 1\,$\sigma$ uncertainties. Hence, we cannot rule out this scenario for dip-A.  \newline

A precessing Jovian planet and a dust clump were the remaining explanations for dip-B. However, the planet hypothesis faces some difficulties, namely the proposed changing inclination with a period of $\sim1411$\,d is based on the proclaimed detection of dip-B on 2018 November 09, which we cannot confirm with our observations of season 2018/2019 in Fig.\,\ref{fig:20182019}. We stress that the presumable planet was never confirmed by $RV$ detections, but a planet scenario was just consistent with $RV$ non-detections. Furthermore, dip-B seems to be in 2015/2016 typically twice as deep in comparison to other seasons from our follow-up observations. This feature makes a precessing planet even more questionable. Testing the dust clump hypothesis with the derived $R_{\text{V}}=5.3$, $f_{\text{core}}=0.003$ and $f_{\text{halo}} \tau_{\text{V}}=0.0027$ from \cite{tanimoto}, shows that these values fit with our observed depths of $\delta_{\text{dip-B}}=0.007\pm0.001$ and $\delta_{\text{dip-B}}0.008\pm0.001$ for season 2014/2015 and 2017/2018, respectively within 2\,$\sigma$. However, they are slightly outside the 3\,$\sigma$ interval for the average depth $\delta_{\text{dip-B}}=0.018\pm0.004$ in season 2015/2016 in the $R$-band.  \newline

\begin{figure}
\includegraphics[width=0.99\columnwidth]{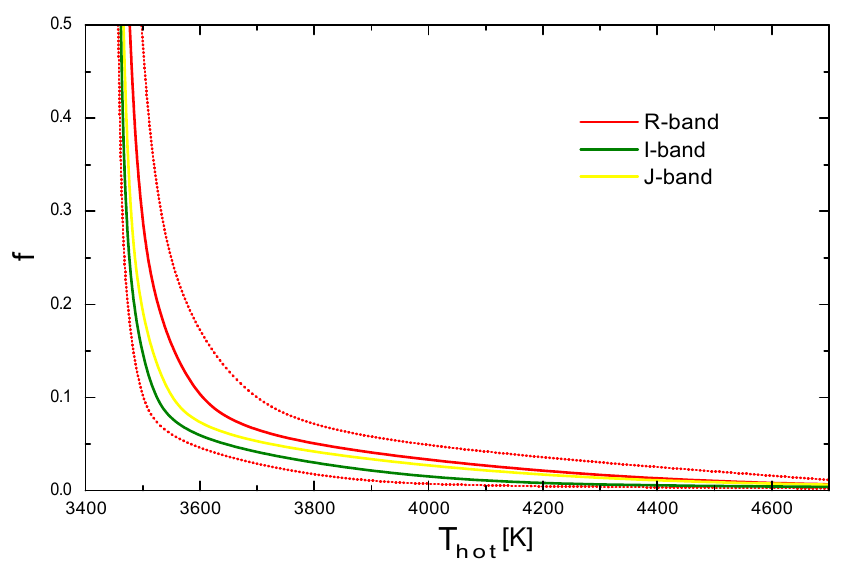}
\caption{Contour map for dip-C's recorded depths of the dimming events. We show the possible combinations of filling factor $f$ and temperature $T_{\text{hot}}$ for an accretion hotspot in different filters. The red dotted lines are the borders of the $3\,\sigma$ intervals for the $R$-band.}
\label{fig:hotspot}
\end{figure}

Dip-C is considered to be either an accretion hotspot or also a circumstellar dust clump. The rotational axis of the star has to be inclined for the hotspot scenario, so that the accreating hotspot, which is brighter than the typical surface area of CVSO\,30, is not visible for the observer during the fading event. The flux depth of a fading event at any wavelength, can therefore be described as
\begin{equation}\label{Tiefehotspot}
\delta_{\text{hot}}(\lambda)=\dfrac{f[B_{\lambda}(T_{\text{hot}})-B_{\lambda}(T_{\star})]}{(1-f)B_{\lambda}(T_{\star})+fB_{\lambda}(T_{\text{hot}})},
\end{equation}
where $f$ is the filling factor and $T_{\text{hot}}$ the temperature of the hotspot \citep{tanimoto}. $T_{\star}$ represents the effective temperature of the star and $B_{\lambda}$ is the brightness for blackbody radiation at a particular wavelength. We use the variables $f$ and $T_{\text{hot}}$ to create the contour map for our observed dimming events of dip-C and those given in \cite{tanimoto}. The average depths in all filters are consistent with each other within $3\,\sigma$ uncertainties as illustrated in Fig.\,\ref{fig:hotspot}. On the other hand, the average depth of  $\delta_{\text{dip-C}}=0.028\pm0.006$ in the $R$-band also satisfies the values $R_{\text{V}}=5.3$, $f_{\text{core}}=0.0065$ and $f_{\text{halo}} \tau_{\text{V}}=0.009$ for a possible circumstellar dust clump within 3\,$\sigma$. Hence, we can neither exclude the hotspot nor the dust clump scenario for dip-C based on the $R$-band photometry.

We have additionally checked if dip-C was detected before 2018 and could be a secondary eclipse of dip-B. The light curves in \cite{raetz} sufficiently cover multiple times the phase of $\sim0.5$ between 2011 and 2013. During this span of time no fading event of dip-C was detected and therefore, we can exclude this hypothesis.

\subsection{Comparison with cluster members}\label{comp}

We used the recently published data from the Early Data Release 3 of the ESA-\textit{Gaia} Mission \citep[\textit{Gaia}\,EDR3,][]{gaiaedr3} to identify members of the 25\,Ori cluster and compare them to CVSO\,30. Therefore, we searched around its prominent member, namely the star 25\,Ori, within a radius of 61\,arcmin for cluster members based on parallax ($\pi$) and proper motion ($\mu$). This search radius takes into account the assumption, that stellar clusters have typical radii up to 5\,pc \citep{Unsoeld2005}, which corresponds to $\sim51$\,arcmin based on the parallax value of 2.9321\,mas for the star 25\,Ori in \textit{Gaia}\,EDR3. We added further 10\,arcmin to this radius, in order not to miss a potential cluster member.

Thereby only sources with significant detected parallaxes and proper motions $(\frac{\pi}{\sigma(\pi)}\geq3, \frac{\mu}{\sigma(\mu)}\geq3)$ were taken into account. The cluster shows an accumulation at $\pi\sim2.9$\,mas within the cumulative distribution function and a common proper motion of $\mu_{\text{RA}}\sim1.4$\,mas/yr in right ascension, but no significant movement towards declination $\mu_{\text{Dec}}\sim0$\,mas/yr. We used at next only stars within sufficiently large intervals around these measurements and performed sigma clipping to identify the most probable cluster members. In total, 239 objects were identified which exhibit on average $\pi=(2.8775\pm0.0718)$\,mas and $\mu_{\text{RA}}=(1.414\pm0.241)$\,mas/yr, while no significant proper motion in declination ($\mu_{\text{Dec}}=-0.256\pm0.580$\,mas/yr) is detectable.

\begin{figure*}
	\includegraphics[width=\textwidth]{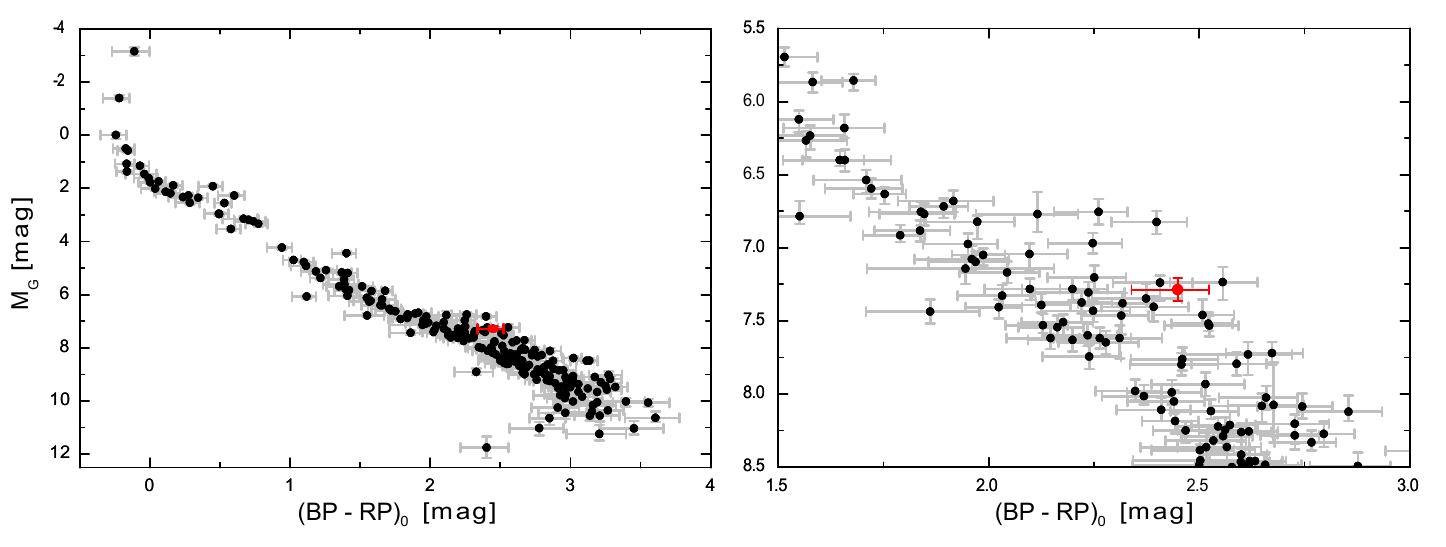}
	\caption{Colour-magnitude diagram of the 25\,Ori cluster with CVSO\,30 marked as red dot.}
	\label{fig:HRD}
\end{figure*}

Based on the photometry of \textit{Gaia}\,EDR3, together with distances from \cite{bailer-jones} and interstellar extinction from the dust maps of \cite{green}, we create the colour-magnitude diagram (CMD) of the 25\,Ori cluster as illustrated in Fig.\,\ref{fig:HRD}. The extinction values were converted into the required pass bands using the relations of \cite{wang}.

Here, we can confirm the results of \cite{koen2020} and \cite{bouma} that CVSO\,30 is about 0.75\,mag brighter than the typical cluster member with comparable colour. This is an indication that CVSO\,30 can be a binary consisting of stellar components with comparable brightness. In this case, we have to modify $M_{\text{G}}$ from Table\,\ref{tab:stardata} into $\tilde{M}_{\text{G}}=8.039_{-0.078}^{+0.075}$\,mag for a single star and following the above procedure, the resulting radius of one stellar component is $\tilde{R}=1.19_{-0.11}^{+0.10}$\,R$_{\odot}$.

\begin{figure}
	\centering
\includegraphics[width=\columnwidth]{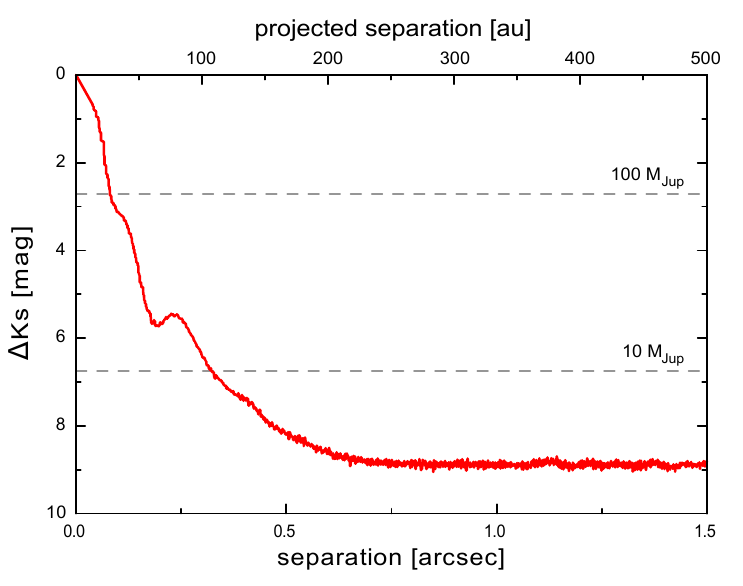}
\caption{The reached detection limit (signal-to-noise ratio $=3$) of the fully reduced NACO $K$s-band image of CVSO\,30, which is presented in Fig.\,\ref{fig:Ks}. The dashed horizontal lines show the expected $K$s-band magnitude differences for companions with 100\,M$_{\text{Jup}}$ and 10\,M$_{\text{Jup}}$ for an 8\,Myr old system, according to models from \protect\cite{baraffe}.}
\label{fig:detect}
\end{figure}

\begin{figure}
	\centering
\includegraphics[width=\columnwidth]{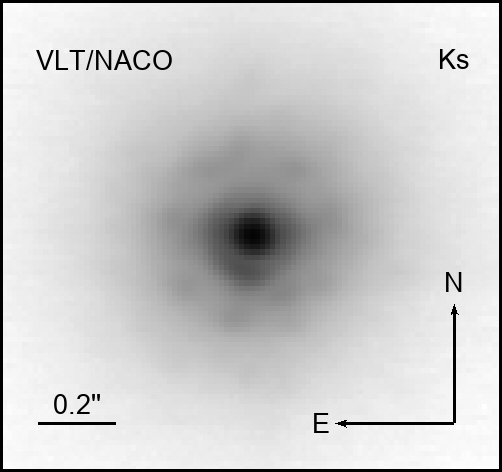}
\caption{Direct imaging with adaptive optics of CVSO\,30.}
\label{fig:Ks}
\end{figure}

If CVSO\,30 is a binary, we might see it either nearly perpendicular to its orbital plane and/or it is a long periodic one, due to the fact that this object shows no line change in radial velocity \citep{vaneyken,ciardi,kounkel}. We used high-resolution direct imaging data with  adaptive optics of CVSO\,30 from the ESO archive to estimate an upper limit on the possible separation, if CVSO\,30 consists of two equally bright stars. The target was observed with NACO at ESO's VLT on December 03 2012 in jitter mode, using a jitter-width of 4\,arcsec, and the data were presented first in \cite{schmidt}. According to the ESO ambient conditions database\footnote{\href{http://archive.eso.org/cms/eso-data/ambient-conditions.html}{\texttt{www.archive.eso.org/cms/eso-data/ambient-conditions.html}}} the average DIMM seeing was $0.67\pm0.03$\,arcsec and the average coherence time of the atmospheric fluctuations was $4.5\pm0.2$\,ms during the $K$s-band observations.
The recorded data contain 15 cubes, each consisting of 4 images with an individual integration time of 15\,s. The frames were flatfielded with internal lamp flats, using the software package ESO ECLIPSE\footnote{\url{https://www.eso.org/sci/software/eclipse/}} \citep{eclipse1997}. We show the reached detection limit for the $K$s-band image of CVSO\,30 in Fig.\,\ref{fig:detect}. The PSF of the star does not exhibit significant elongation in any direction, as illustrated in  Fig.\,\ref{fig:Ks}. It exhibits a FWHM of $6.4\pm0.6\,\text{px}$. Adopting the pixel scale $13.265\pm0.041$\,mas/px from \cite{schmidt}, this corresponds to $85\pm8$\,mas. The diffraction limit of the 8.2\,m VLT in the $K$s-band is about $68$\,mas.\\
In order to test, at which separation an equally bright companion can be detected, we shifted the fully reduced image pixel by pixel, averaged it with the original frame and fit a two dimensional Gaussian function with ESO-MIDAS \citep{esomidas}. The artificial PSF becomes clearly elongated for equal bright sources with an angular separation larger $40$\,mas. This is in good agreement with \cite{Mugrauer2015}, where they have detected a close binary companion of the exoplanet host star HD\,142245 with NACO in the $K$s-band. That binary shows a clearly elongated PSF and an average separation of about $40$\,mas of its components.\\
Therefore, in our case a possible equal bright binary would have to be within a separation of 40\,mas in order to have not been detected within the observations of CVSO\,30. Based on this separation and a distance of $334_{-3}^{+4}$\,pc \citep{bailer-jones}, we expect an upper limit of the orbital period of about 50\,yr for the CVSO\,30 system, assuming a total mass of 1\,M$_{\sun}$ (corresponding to two times the mass given in Table\,\ref{tab:stardata}). Additional observations are necessary to further constrain this upper limit, such as high-resolution follow-up spectroscopy or interferometric observations of CVSO\,30 in the upcoming decades, as proposed by \cite{koen2020}.

Otherwise, it might be also possible that CVSO\,30 is redder ($\sim0.25$\,mag) than the typical cluster member. For a given mass, increasing metallicity shifts the stellar flux from the visual wavelengths range towards the infrared \citep{Bonfils2005,Mann2015}. Therefore, we searched for information regarding the metallicity of the 25\,Ori members. The catalog of \cite{anders} contains 39 cluster stars, which are located in the CMD within CVSO\,30's $3\,\sigma$ uncertainties. These stars have a median metallicity of $[\text{Fe/H}]=(0.40\pm0.15)$\,dex, which is consistent with CVSO\,30 (see Table\,\ref{tab:stardata}) within the standard deviation. Therefore, metallicity cannot explain the offset of CVSO\,30 within the CMD. If CVSO\,30 is actually redder than the other member stars, it needs an additional source, which contributes to the flux in the red wavelengths. This could be an orbiting co-rotating glowing cloud, that shows significant H$_{\alpha}$ emission. We consider the possibility of this scenario in section\,\ref{discussion}.

\section{Discussion and conclusion}\label{discussion}

In this article we presented our follow-up photometric observations of the controversial discussed star CVSO\,30, which was intensively monitored and analyzed during the last decade, but still misses a clear explanation for its periodically dimming events. Our original YETI monitoring campaign of CVSO\,30 started in 2010 \citep{raetz} and this is the continuation, which yields observations since fall 2014 that focused on the predicted time slots of the fading events.

We characterized CVSO\,30 with data of the $Gaia$ mission and catalogs. Our derived value of $1.69\pm0.16$\,R$_{\sun}$ for its radius lies above those given in \cite{briceno} (1.39\,R$_{\sun}$) and \cite{koen2020} (1.41\,R$_{\sun}$), but is consistent with them within $2\,\sigma$. In contrast, our radius does not fit the $0.45\pm0.18$\,R$_{\sun}$ from \cite{tanimoto}.

The different dimming events, dip-A, dip-B and dip-C, reported in \cite{tanimoto} can be confirmed with our $R$-band observations. All three dips seem to have the same period but are phase shifted as illustrated in Fig.\,\ref{fig:OC}. Dip-A was detected by us in all nights if the observing window included the phase of $-0.1$ according to the ephemeris in Eqn.\,\ref{ephem}, while dip-B, e.g., was not present on January 5, 2017. Dip-A and dip-B were detected last in November 2017 and since then no more, as presented in the following light curves in appendix\,\ref{appLC}. That contradicts the detection of dip-B on 09 November 2018 by \cite{tanimoto}, because we have recorded the immediately following epoch without any fading event. Dip-A showed sometimes a "v"-shaped profile and then in other epochs a "u"-profile, as illustrated in Fig.\,\ref{fig:phaseAB}. The same also applies to dip-B.

Dip-C was found first by \cite{tanimoto} in autumn 2018. The first successful observation in our data set was in October 2020. Five further detections followed until February 2021. All light curves of dip-C yield "u"-shape like minima.

For all three dips no clear trend was detectable in depth or duration, taking into account the entire period of the follow-up observations.

A circumstellar dust clump, consisting of an opaque core and a optically thin halo, cannot be excluded as cause for all dimming events based on their depths in the $R$-band and infrared. The existence of these clumps could be temporally limited and therefore explain the disappearance of dip-A and dip-B. As a result thereof, additional observations of dip-C in upcoming seasons are necessary to investigate whether this dip might vanish too.

However, such a theoretical clump would orbit around CVSO\,30 at a distance of $(9.1\pm0.2)\cdot10^{-3}$\,au and have to face temperatures of $2268_{-141}^{+161}$\,K. The expected condition at this location is above the sublimation temperatures for olivine, pyroxene, obsidian, iron, ice and carbon, which were derived based on \cite{Kobayashi2011}. Therefore, the clump will probably consists of gas rather than dust, which could be problematic for the interpretation by \cite{Grosson2021}. A gas cloud would fit better to a weak-line T Tauri star, because dust would have resulted into an infrared excess. Weak-line T Tauri stars have no or only an optically thin disc. Building on this consideration, the flux dips happen when the glowing cloud orbits behind the star \citep{stauffer2017,jardine}. A system with gas arranged in a circumstellar clumpy torus, could result from stellar winds at locations around the star where magnetic and gravitational forces are balanced \citep{jardine, Cameron1989}. In contrast, \cite{David2017} have presented a scenario, where a cloud, containing a minor amount of dust within the gas, around the young M\,2.5 dwarf star RIK-210 could cause dimming up to $\sim20\%$. This star shows variable dimming events with a period of $\sim5.7$\,d and the co-rotating orbit lies, as a result thereof, outside the sublimation distance. In the case of RIK-210 the dips would occur when the star is eclipsed by a dusty gas cloud.\\

Our case also involves co-rotating material, but we likely see secondary eclipses of light emitting gas. In this case, co-rotating plasma, which emits in a specific wavelength, causes the dimming events when it orbits behind the star and a reduced flux is measured \citep{palumbo}. This orbiting plasma would also result in changes of the H$_{\alpha}$ emission line profile via Rositter-McLaughlin effect (RME) when it moves in front of the star's disc as stated by \cite{palumbo}. This feature was not observed in the H$_{\alpha}$ measurements by \cite{yu}, who have recorded a set of spectra from CVSO\,30 on 2013 December 12, which include the time of a photometric fading event. Their observations started/ended about 2\,h before/after the minimum light and cover the timespan when the postulated glowing cloud would orbit behind the star. Therefore, a non-detection of the RME is well explainable with this scenario.
Further H$_{\alpha}$ observations were presented by \cite{Krull2016}. They have measured significant H$_{\alpha}$ excess emission within the spectra of CVSO\,30, that changes its $RV$. The changing $RV$ fits with the orbital period of the "companion". However, the measured $RV$s of the excess emission are often shifted in comparison to the predicted velocities \citep[see Fig.\,9 in][]{Krull2016} based on the ephemeris by \cite{vaneyken}. Furthermore, the strength of the detected H$_{\alpha}$ excess is too large (about $70\%-80\%$ of the stellar equivalent width) to be caused by a single planet and needs an extended additional luminous volume, which surrounds it \citep{Krull2016}. All these observed features fit well with a co-rotating glowing cloud, which significantly emits flux in the optical $R$-band. \cite{David2017} stated that those clouds, consisting of partially ionized gas, can cause dimming events up to a few percents when they move behind the star and the cloud's glowing could be the result of Paschen-continuum bound-free emission. The existence of partially ionized gas which emits in H$_{\alpha}$ is quite possible even at relatively low temperatures such as $\sim2000$\,K \citep{Rodriguez2015}.

\cite{stauffer2017,stauffer2018,Stauffer2021} had analyzed a sample of photometric variable mid-to-late type M dwarfs in star-forming regions without signs of active accretion. These targets are rapidly rotating weak-line T Tauri stars with photometric periods shorter than one day. A subgroup of this sample are the "stars with persistent flux dips", which additionally show two to four discrete flux dips in their phased light curve, with deepest depths of $2\%-7\%$ and durations ranging between 1\,h and 5\,h. Their depths are largely stable but can suddenly disappear or become significantly weaker. Those changes in depth were observed after the detection of flare-like events. Similar fast rotating young M-dwarfs were found by \cite{zhan}. We detected some flare-like signals within our observations and they occurred after the last significant detection of dip-A and dip-B as illustrated in Fig.\,\ref{fig:OC}. Similar signatures can be found within the presented light curves in Fig.\,2 and Fig.\,5 in \cite{vaneyken}, which contain observations from 26 Dec 2009, 01 Jan 2010, 07 Jan 2010 and 09 Feb 2011. The scenario of \cite{stauffer2017,stauffer2018,Stauffer2021} has the flexibility to explain the dis- and reappearance of the dimming events by changes in the geometry of the cloud without relying on active accretion. CVSO\,30 shows the same characteristics in its light curve as the stars in \cite{stauffer2017,stauffer2018,Stauffer2021} and is also similar to the detected features in the light curve of TIC\,234284556 \citep{palumbo}. Therefore, the dimming events could originate from the currently not well understood process of gas tori around young M dwarfs. Dip-A, -B and -C seem to have the same orbital period, which is why their origin should have also the same distance to the star. This is a further indication for the theory of emitted light from magnetospheric clouds. Furthermore, the measured depth $\Delta m$ in Table\,\ref{tab:transitsETD} would be the result of a secondary transit and consequently
\begin{equation}\label{sektrans}
\Delta m = -2.5\log\left(\frac{F_{\star}}{F_{\star}+F_{\text{cloud}}}\right),
\end{equation}
where $F_{\star}$ represents the stellar flux and $F_{\text{cloud}}$ the flux of the cloud. According to Stefan-Boltzmann law $F_{\star}\equiv R^{2}\cdot T_{\text{eff}}^{4}$ for the single star or $F_{\star}\equiv2\cdot \tilde{R}^{2}\cdot T_{\text{eff}}^{4}$ for the binary scenario, respectively.

We consider it well possible that flares due to reconnection of magnetic field lines and plasma tubes also lead to a reconfiguration of spots and groups of spots on the surface, which potentially affect the optical light curve. A strong X-ray flare in CVSO\,30's light curve is reported in \cite{czesla}, where they have found no significant transit-induced variation within the expected time slot. That makes an orbiting planet even more questionable. The bell like shape of the X-ray flare may also be owing to an accretion episode of CVSO\,30. That fits with \cite{yu}, who came to the conclusion that CVSO\,30 may be weakly accreting based on their measured strength and breadth of the H$_{\alpha}$ line profile.

\cite{koen2021} presents a model, where CVSO\,30 is considered to be a binary and the variability is the result of stellar spots. From pure statistics the model can reproduce the detected variations inside the \textit{TESS} photometry, but the assumed filling factors e.g. $f\gtrsim0.5$ for norm-2 models are very large and are based only on measurements in one filter. The detected fading events by \textit{TESS} are dip-C according to Fig.\,\ref{fig:OC}, for which \cite{tanimoto} had already ruled out the cool star spot scenario based on their multi-band photometry. The two different (apparent rotation) periods within the light curve of CVSO\,30 \citep{bouma, koen2020} could also be caused by two spots of different latitudes, if the star (i.e. just one star) rotates differentially. \cite{reinhold2013} and \cite{reinhold2015} showed that the relative shear $\alpha=(P_{\text{max}}-P_{\text{min}})/P_{\text{max}}$ increases with rotation period, by analyzing thousands of stars from the $Kepler$-mission \citep{koch}. That fits with the two short-periodic signals of CVSO\,30, which are close together in time.  \\

To sum up the results of our monitoring campaign, we confirm the detection by \cite{tanimoto} of three phase shifted dimming events in the light curve of CVSO\,30 with our photometric data between 2014 and 2021. Dip-A and dip-B were detected by us last in autumn 2017 and seem to have been vanished since then, while another dip-C was found after that, shifted in phase at about $180^{\circ}$. A Jovian planet as cause for the dimming events is unlikely, because of the colour effects of the transit depths and their disappearance within relatively short timescales. We agree with \cite{bouma} that orbiting clouds of gas at a Keplerian co-rotating radius are the most promising scenario to explain most changes in CVSO\,30's light curve, because it does not need active accretion from a circumstellar disc and a changing shape of the dimming events can result from changes in the cloud's geometry. However, we consider that also stellar spots and at least some accretion seems to be going on CVSO\,30. Nevertheless, further follow-up observations are necessary to find out if more flare-like events occur in the future, right before changes of dip-C can be detected. Furthermore, additional high-resolution spectroscopy and interferometric follow-up observations should be done to test CVSO\,30's multiplicity status as proposed by \cite{koen2020}.

\section*{Acknowledgements}

This work is based on observations obtained with telescopes of the University Observatory Jena, operated by the Astrophysical Institute of the Friedrich-Schiller-Universit\"{a}t Jena. We thank B. Baghdasaryan, N. Belko, S. Buder, M. Dadalauri, M. Geymeier, H. Gilbert, A. Gonzalez, F. Hildebrandt, H. Keppler, O. Lux, S. Masda, P. Protte, J. Trautmann, A. Trepanowski, and S. Schlagenhauf, who have been involved in some observations of this project, obtained at the University Observatory Jena.

This research was partly based on data obtained at the 1.5\,m telescope of the Sierra Nevada Observatory (Spain), which is operated by the Consejo Superior de Investigaciones Cient\'{\i}ficas (CSIC) through the Instituto de Astrof\'{\i}sica de Andaluc\'{\i}a. We thank J.F.Aceituno and V.Casanova for their help with the observations.

This publication is partly based on observations made with ESO Telescopes at the La Silla Paranal Observatory under programme ID 090.C-0448(A).

RB, RN and MM acknowledge the support of the DFG priority program SPP\,1992 "Exploring the Diversity of Extrasolar Planets" in projects NE\,515/58-1 and MU\,2695/27-1.

We acknowledge financial support from the Spanish Agencia Estatal de Investigaci\'{o}n of the Ministerio de Ciencia, Innovaci\'{o}n y Universidades and the ERDF through projects PID2019-109522GB-C52 and AYA2016-79425-C3-3-P, and the Centre of Excellence "Severo Ochoa" award to the Instituto de Astrof\'{\i}sica de Andaluc\'{\i}a (SEV-2017-0709).

We thank R. Errmann for developing and providing the photometry routine "automat.py" and also C. Broeg his program "PHOTOMETRY".

This publication makes use of data products of the SIMBAD and VizieR databases, operated at CDS, Strasbourg, France. We also thank the \textit{Gaia} Data Processing and Analysis Consortium of the European Space Agency (ESA) for processing and providing the data of the \textit{Gaia} mission. We thank the MAST portal for providing the \textit{TESS} photometric results and the MESA Isochrones \& Stellar Tracks website for the bolometric correction tables.

We thank the referee for helpful comments, which improved our manuscript.

\section*{Data Availability}

The data underlying this article are available in this manuscript. The detrended (if applicable) and also the original photometric measurements of all light curves are provided as online supplementary material.

\bibliographystyle{mnras}
\bibliography{bischoff}

\vfill
\newpage
\appendix\section{Light curves and results of the transit fitting}\label{appLC}

\begin{figure*}
\includegraphics[width=\textwidth]{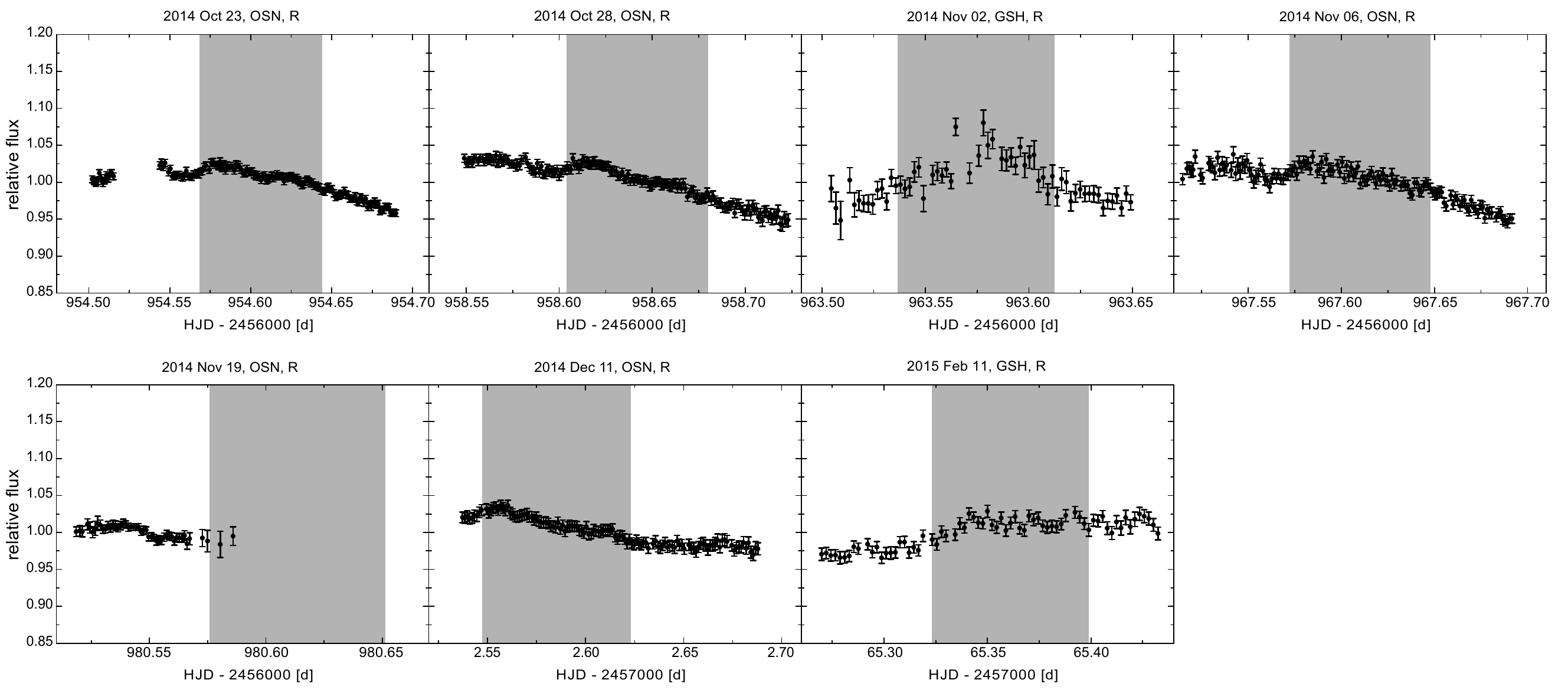}
\caption{All recorded light curves of CVSO\,30 in season 2014/2015. The grey shaded areas show the time slots of the expected fading events, fixed at period and mid-time from \protect\cite{tanimoto}. The value for the duration was taken from \protect\cite{vaneyken} and is fixed in all light curves.}
\label{fig:20142015}
\end{figure*}

\begin{figure*}
	\includegraphics[width=\textwidth]{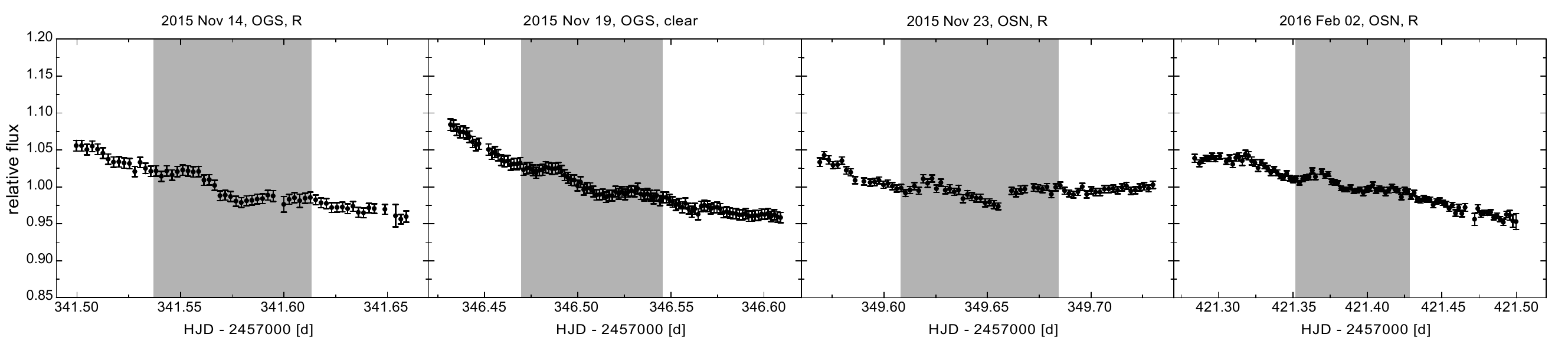}
	\caption{Same as Fig.\,\ref{fig:20142015} but for season 2015/2016.}
	\label{fig:20152016}
\end{figure*}

\begin{figure*}
	\includegraphics[width=\textwidth]{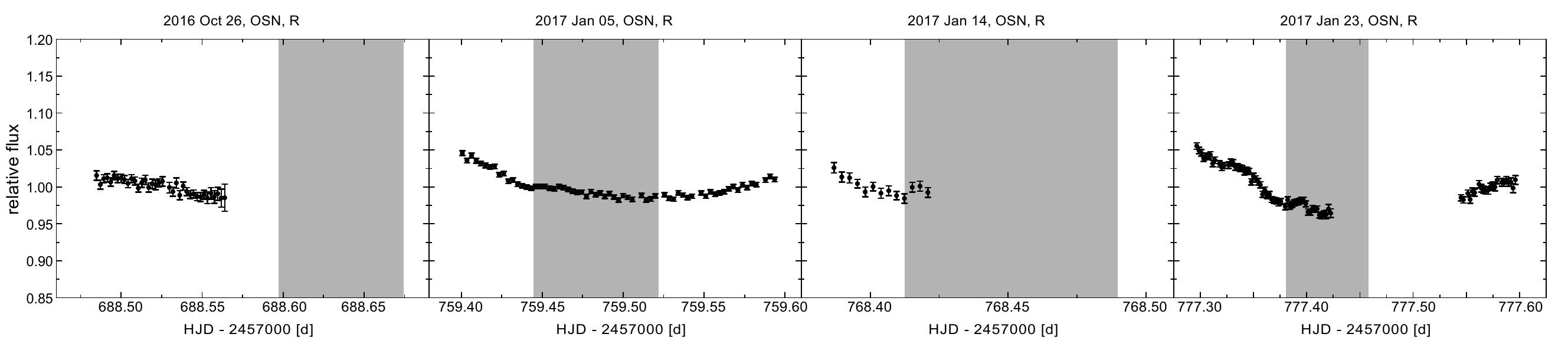}
	\caption{Same as Fig.\,\ref{fig:20142015} but for season 2016/2017.}
	\label{fig:20162017}
\end{figure*}

\begin{figure*}
	\includegraphics[width=\textwidth]{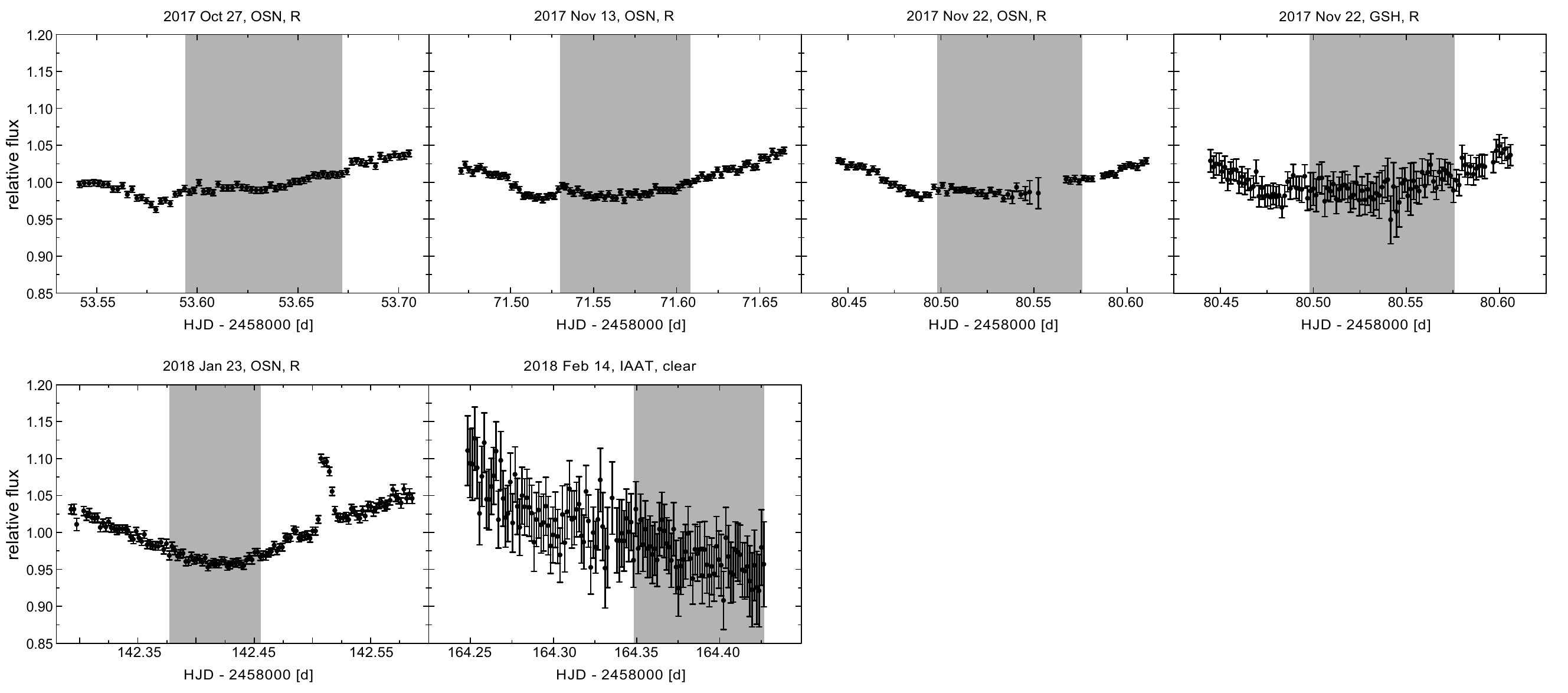}
	\caption{Same as Fig.\,\ref{fig:20142015} but for season 2017/2018.}
	\label{fig:20172018}
\end{figure*}

\begin{figure*}
	\includegraphics[width=\textwidth, height=13.1cm]{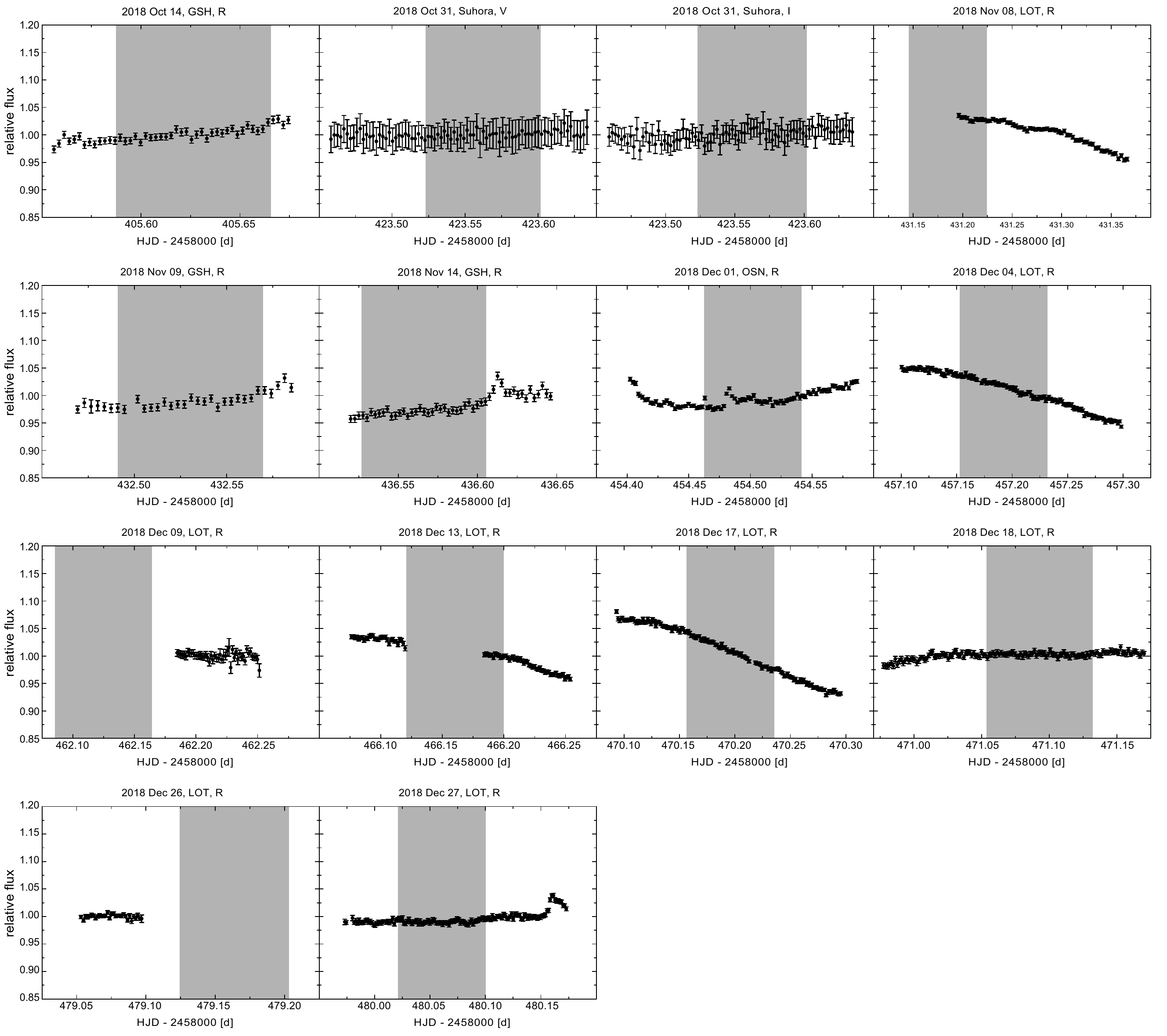}
	\caption{Same as Fig.\,\ref{fig:20142015} but for season 2018/2019.}
	\label{fig:20182019}
\end{figure*}

\begin{figure*}
	\includegraphics[width=\textwidth]{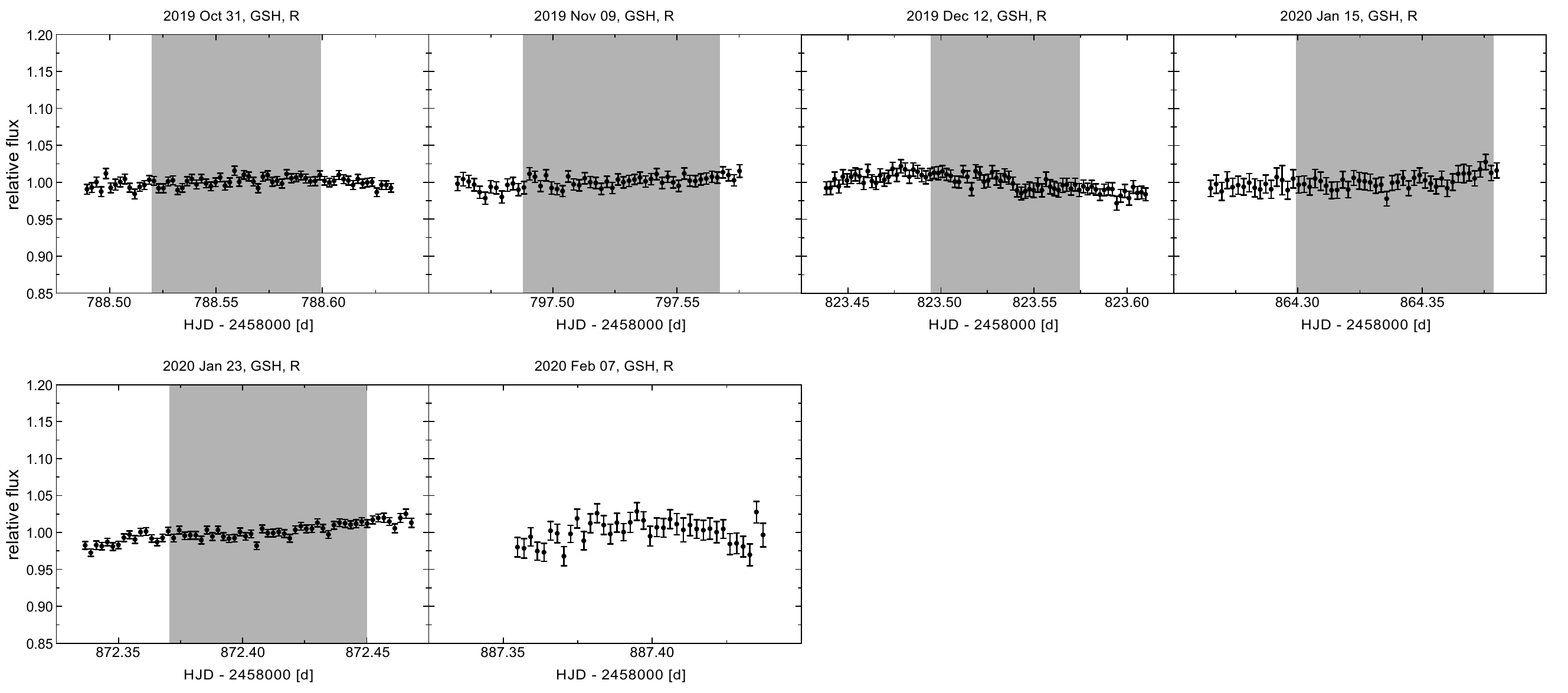}
	\caption{Same as Fig.\,\ref{fig:20142015} but for season 2019/2020.}
	\label{fig:20192020}
\end{figure*}

\begin{figure*}
	\includegraphics[width=\textwidth]{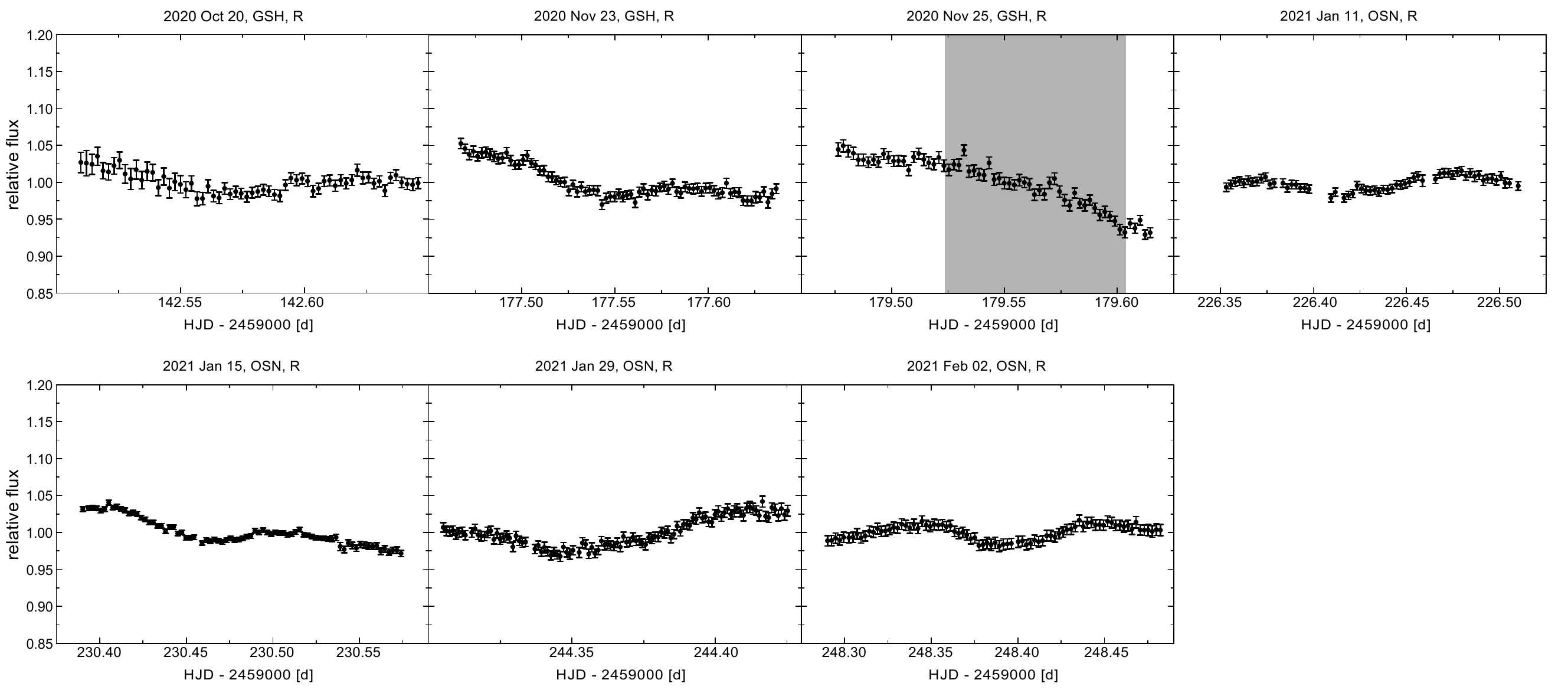}
	\caption{Same as Fig.\,\ref{fig:20142015} but for season 2020/2021.}
	\label{fig:20202021}
\end{figure*}

\begin{table*}
\centering
\caption{Results of the transit fitting for the 29 significant detected fading events of CVSO\,30 carried out with \texttt{ETD}. We list for each epoch the date (start of observation), transit center time ($T_{c}$), duration ($T_{14}$), depth ($\Delta m$), and orbit phase according to the ephemeris from \protect\cite{tanimoto}.}
\label{tab:transitsETD}
\begin{tabular}{clcccc}
\hline
Epoch & Date	&	$T_{c}$ $[\text{BJD}_{\text{TDB}}]$  & $T_{14}$ $[\text{min}]$ 	& $\Delta m$ $[\text{mmag}]$ & orbit phase \\
\hline
Dip-A &&&&&\\
3146      & 2014 Oct 23		&	$2456954.5611\pm0.0007$ 	 & $\,\,\,43.6\pm4.8$ 		& $16.5\pm1.7$		& $-0.103\pm0.006$ \\
3155      & 2014 Oct 28		&	$2456958.5946\pm0.0011$ 	 & $\,\,\,64.7\pm6.3$ 		& $18.5\pm2.0$		& $-0.107\pm0.007$ \\	
3175      & 2014 Nov 06		&	$2456967.5630\pm0.0011$ 	 & $\,\,\,53.3\pm7.1$ 		& $22.5\pm2.8$		& $-0.107\pm0.007$ \\		
3393      & 2015 Feb 11		&	$2457065.3098\pm0.0021$ 	 & $\,\,\,113.4\pm14.6$ 	& $29.1\pm4.6$		& $-0.116\pm0.008$ \\		
4009      & 2015 Nov 14		&	$2457341.5318\pm0.0015$ 	 & $\,\,\,\,\,\,77.6\pm10.2$& $14.0\pm2.0$		& $-0.098\pm0.008$ \\		
4020      & 2015 Nov 19$^{a}$&	$2457346.4643\pm0.0009$  	 & $\,\,\,87.1\pm6.1$ 		& $17.8\pm1.8$		& $-0.098\pm0.007$ \\		
4027      & 2015 Nov 23		&	$2457349.6013\pm0.0014$ 	 & $\,\,\,82.0\pm9.3$ 		& $18.2\pm2.4$		& $-0.102\pm0.008$ \\		
4187      & 2016 Feb 02		&	$2457421.3482\pm0.0012$ 	 & $\,\,\,70.8\pm7.8$ 		& $17.1\pm1.7$		& $-0.095\pm0.008$ \\
4941      & 2017 Jan 05		&	$2457759.4385\pm0.0012$ 	 & $\,\,\,55.4\pm7.2$ 		& $11.3\pm1.4$		& $-0.101\pm0.008$ \\
4981      & 2017 Jan 23		&	$2457777.3770\pm0.0011$ 	 & $\,\,\,89.4\pm7.3$ 		& $24.7\pm1.7$		& $-0.096\pm0.008$ \\		
5597      & 2017 Oct 27		&	$2458053.5797\pm0.0008$ 	 & $\,\,\,44.9\pm5.2$ 		& $22.1\pm2.4$		& $-0.121\pm0.009$ \\
5637      & 2017 Nov 13		&	$2458071.5155\pm0.0008$ 	 & $\,\,\,54.3\pm5.1$ 		& $22.1\pm2.0$		& $-0.121\pm0.009$ \\
5657      & 2017 Nov 22$^{b}$&	$2458080.4855\pm0.0009$	 	 & $\,\,\,60.9\pm5.9$ 		& $17.5\pm1.7$		& $-0.117\pm0.009$ \\
5657      & 2017 Nov 22$^{c}$&	$2458080.4781\pm0.0016$ 	 & $\,\,\,\,\,\,35.0\pm10.0$& $20.7\pm5.4$		& $-0.133\pm0.009$ \\
		&&&&&\\
Dip-B	&&&&&\\ 		
3146      & 2014 Oct 23		&	$2456954.6072\pm0.0013$		 & $\,\,\,37.6\pm7.6$ 		& $\,\,\,7.9\pm1.5$	& $\,\,\,\,0.000\pm0.007$ \\
3155      & 2014 Oct 28		&	$2456958.6410\pm0.0015$		 & $\,\,\,52.5\pm8.8$ 		& $\,\,\,9.0\pm1.5$	& $-0.004\pm0.007$ \\
3253      & 2014 Dec 11		&	$2457002.5839\pm0.0015$ 	 & $\,\,\,38.2\pm8.6$ 		& $\,\,\,7.1\pm1.5$	& $-0.005\pm0.007$ \\
4009      & 2015 Nov 14		&	$2457341.5786\pm0.0007$ 	 & $\,\,\,57.2\pm4.6$ 		& $25.2\pm1.9$		& $\,\,\,\,0.006\pm0.007$ \\
4020      & 2015 Nov 19$^{a}$&	$2457346.5101\pm0.0007$  	 & $\,\,\,58.9\pm6.1$ 		& $16.7\pm1.2$		& $\,\,\,\,0.004\pm0.007$ \\
4027      & 2015 Nov 23		&	$2457349.6486\pm0.0010$ 	 & $\,\,\,62.4\pm5.7$ 		& $20.9\pm1.8$		& $\,\,\,\,0.004\pm0.007$ \\
4187      & 2016 Feb 02		&	$2457421.3892\pm0.0012$ 	 & $\,\,\,55.5\pm7.1$ 		& $14.8\pm1.7$		& $-0.004\pm0.008$ \\
5597      & 2017 Oct 27		&	$2458053.6392\pm0.0021$ 	 & $\,\,\,\,\,\,57.5\pm12.7$& $\,\,\,7.7\pm1.6$	& $\,\,\,\,0.012\pm0.010$ \\
5637      & 2017 Nov 13		&	$2458071.5703\pm0.0025$ 	 & $\,\,\,122.1\pm11.0$ 	& $\,\,\,8.9\pm1.1$	& $\,\,\,\,0.001\pm0.010$ \\
		&&&&&\\
Dip-C	&&&&&\\
8026      & 2020 Oct 20		&	$2459142.5694\pm0.0024$ 	 & $\,\,\,\,\,\,92.7\pm14.5$& $17.2\pm2.6$		& $-0.505\pm0.013$ \\
8104      & 2020 Nov 23		&	$2459177.5427\pm0.0012$ 	 & $119.2\pm7.0$ 			& $30.0\pm1.7$		& $-0.509\pm0.012$ \\
8213      & 2021 Jan 11		&	$2459226.4220\pm0.0009$ 	 & $148.7\pm4.5$ 			& $33.4\pm1.2$		& $-0.501\pm0.012$ \\
8222      & 2021 Jan 15		&	$2459230.4588\pm0.0007$ 	 & $138.9\pm4.0$ 			& $33.5\pm1.0$		& $-0.498\pm0.012$ \\
8253      & 2021 Jan 29		&	$2459244.3623\pm0.0010$ 	 & $129.4\pm5.4$ 			& $30.4\pm1.4$		& $-0.491\pm0.012$ \\
8262      & 2021 Feb 02		&	$2459248.3965\pm0.0006$ 	 & $134.9\pm3.3$ 			& $37.2\pm0.9$		& $-0.494\pm0.012$ \\
\hline
\end{tabular}
\newline
\begin{flushleft}
$^{a}$ clear filter, $^{b}$ OSN, $^{c}$ GSH\\
\end{flushleft}
\end{table*}

\bsp
\label{lastpage}
\end{document}